\documentclass[runningheads]{svmult}

\usepackage{makeidx}   % allows index generation
\usepackage{graphicx}  % standard LaTeX graphics tool
\usepackage{subeqnar}  % subnumbers individual equations
\usepackage{multicol}  % used for the two-column index
\usepackage{physprbb}  % modified textarea for proceedings,
\makeindex             % used for the subject index
\newcommand {\arcsec}{\mbox{$^{\prime\prime}$}}
\newcommand {\asp}{\mbox{$.\!\!^{\prime\prime}$}}
\newcommand {\grd}{\mbox{$^{\circ}$}}

\begin{document}
%
%\title*{A close view on a protoplanetary disk}
\title*{A close view on the protoplanetary disk in the Bok globule CB\,26}
\toctitle{Focusing of a Parallel Beam to Form a Point
\protect\newline in the Particle Deflection Plane}
% allows explicit linebreak for the table of content
%
%
\titlerunning{A protoplanetary disk in CB\,26}
% allows abbreviation of title, if the full title is too long
% to fit in the running head
%
\author{R. Launhardt \inst{1}
\and B. Stecklum\inst{2}
\and A.I. Sargent\inst{1}}
\authorrunning{R. Launhardt et al.}
% if there are more than two authors,
% please abbreviate author list for running head
%
%
\institute{Astronomy Department, California Institute of Technology, 
    Pasadena, CA 91125
\and Th\"uringer Landessternwarte Tautenburg, 07778 Tautenburg, Germany}

\maketitle              % typesets the title of the contribution

\begin{abstract}
We present new sub-arcsecond-resolution near-infrared polarimetric imaging 
and millimetre interferometry data on the circumstellar disk system in the 
Bok globule CB26. The data imply the presence of a $M\ge 0.01\,$M$_{\odot}$\ 
edge-on-seen disk of $> 400$\,AU in diameter, being in Keplerian rotation 
around a young $\sim$0.35\,M$_{\odot}$\ star. 
The mm dust emission from the inner 200\,AU is highly optically thick, but 
the outer parts are optically thin and made of small dust grains. 
Planetesimal growth in the inner disk could neither be comfirmed nor excluded. 
%The outer optically thin disk is strongly warped. 
We argue that the CB\,26 disk is a very young protoplanetary disk 
and show that it is comparable to the early solar system.

\end{abstract}

%%%%%%%%%%%%%%%%%%%%%%%%%%%%%%%%%%%%%%%%%%%%%%%%%

%\section{From accretion disks to planetary systems}
\section{Introduction}
Observations of protostellar systems and their prominent jets and outflows 
suggest that accretion disks start to form very early during the main accretion 
phase. These disks live much longer than the central 
protostar needs to build up most of its mass. When the initial protostellar 
core is dispersed by accretion and outflows, the central star 
still accretes matter at low rates from the surrounding disk. 
Typical disk life times around low-mass stars were shown to be at least 
$10^7$\,yrs and their masses (typically 0.01-0.1\,M$_{\odot}$) do not seem 
to decrease considerably during this time \cite{bec90}.
Theoretical and laboratory studies show that the timescale for grain growth 
and planetesimal formation is shorter than the typical disk life time. 
Such disks should soon evolve into protoplanetary disks. 
Although the direct detection of planetary-mass bodies in such disks 
will be extremely difficult, indirect effects of larger bodies, such as 
a change in the dust opacity spectrum, should be observable with 
current or near-future techniques.

%%%%%%%%%%%%%%%%%%%%%%%%%%%%%%%%%%%%%%%%

%\section{The Bok globule CB\,26}
%
CB\,26 (L\,1439) is a small ($d\sim 0.15$\,pc), 
slightly cometary-shaped Bok globule 
%located $\sim$10\grd\ north from the Taurus/Auriga dark cloud 
at a distance of $\sim 140$\,pc \cite{lsz01}.
Located at the south-east rim of the globule is a small bipolar 
near-infrared (NIR) nebula.
% with a dark extinction lane at its center. 
%High angular resolution NIR imaging polarimetry of the nebula revealed 
%a polarization pattern which could be modeled in terms of an almost 
%edge-on seen circumstellar disk and a thin envelope \cite{ste01}. 
The central star is not visible, even at 2.2\,$\mu$m.  
The spectral energy distribution 
together with the bolometric luminosity of $\ge 0.7$\,L$_{\odot}$\  
suggest the presence of a low-mass Class I YSO in CB\,26. 
We observed strong submm/mm dust continuum emission from this source 
showing a thin extended envelope with an unresolved condensation 
at 10\arcsec\ resolution \cite{lh97}, \cite{lsz01}, \cite{hen01}. 
%From submm continuum imaging polarimetry we derived a well-ordered 
%magnetic field in the envelope with $\langle B\rangle=(90\pm50)$\,$\mu$G
%(\cite{lh97}, \cite{lsz01}, \cite{hen01}).
No molecular outflow has been detected yet.

%%%%%%%%%%%%%%%%%%%%%%%%%%%%%%%%%%%%%%%%

\section{Implication of a circumstellar disk from NIR data}

\begin{figure}[htb]
\begin{center}
\includegraphics[width=0.8\textwidth]{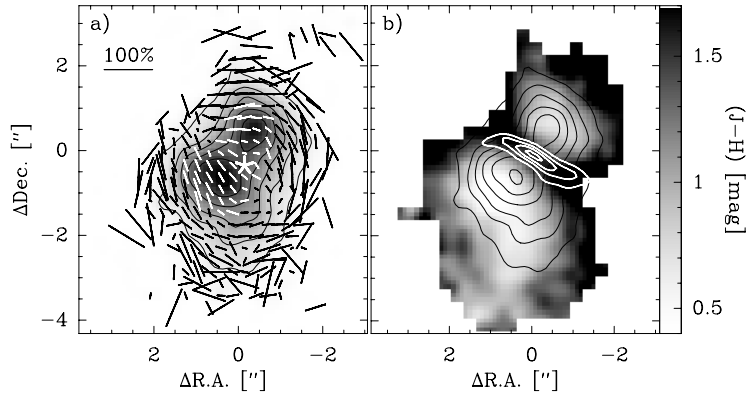}
\end{center}
\caption{\label{fig_jmh}
   {\bf a)} K-band image of the CB\,26 NIR nebula with polarization 
    vectors superimposed. The white star marks the center of the 
    polarization pattern, i.e., the location of the illuminating 
    source.~~~
   {\bf b)} Grey-scale image: J\,--\,H color map of the 
   NIR reflection nebula. Darker regions represent higher extinction. 
   The outer boundary is due to an intensity cut-off level. 
   Overlayed are the contours of the K-band emission from 
   the reflection nebula (black) and of the 1.3\,mm dust continuum emission 
   from the circumstellar disk.
   The K-band image and color map are from \cite{ste01}.
} 
\end{figure}

On JHK near-infrared images obtained with the MAGIC camera at the 
Calar Alto 3.5-m telescope, we found a small bipolar reflection nebula 
in the Bok globule CB\,26 within the error ellipse of the IRAS source 
04559+5200 \cite{ste01}. 
Subsequent NIR imaging polarimetry 
%at higher angular resolution 
confirmed the bipolar structure of this source. 
The two lobes are separated by an extinction lane which is most obvious 
in the J--H color map (Fig. \ref{fig_jmh}b).
Very high polarization degrees were detected in the lobes, 
presumably caused by single scattering at small dust grains. 
The orientation of the polarization vectors corresponds to a system 
consisting of a young star surrounded by both a circumstellar 
disk and a thin envelope. The polarization pattern indicates that the
disk is seen almost edge-on causing the band of enhanced
extinction in between the scattering lobes. 
At the very center, the polarization vectors are aligned linearly. 
This could be either due to photons scattered
back from the envelope onto the disk or because of multiple
scattering in the outer disk regions.
The location of the illuminating source 
was derived to an accuracy of 0\asp3 
by minimizing the mean square scalar product between 
polarization vectors in the lobes which are probably due to single 
scattering and their corresponding normalized radius vectors. 
The central source is located behind 
the extinction lane, i.e., at the center of the disk 
(Fig. \ref{fig_jmh}a).

%%%%%%%%%%%%%%%%%%%%%%%%%%%%%%%%%%%%%%%%

\section{Direct observations of dust and gas}

%%%%%%%%%%%%%%%%%%%%%%%%%%%%%%%%%%%%%%%%

%\subsection{Dust continuum emission}

\begin{figure}[htb]
\begin{center}
\includegraphics[width=.7\textwidth]{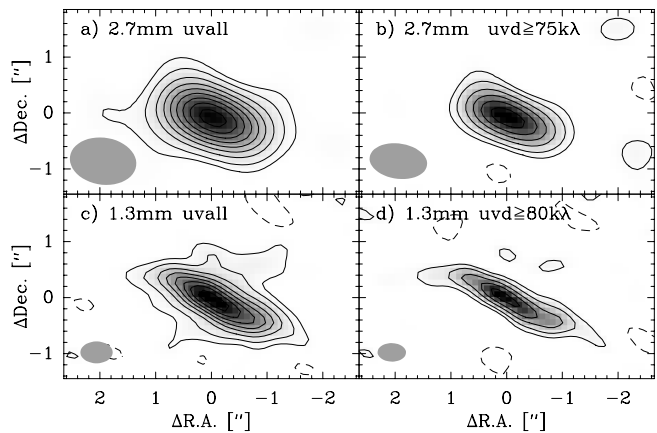}
\end{center}
\caption{\label{fig_mmcres}
   Dust continuum emission from the CB\,26 disk: OVRO results (from \cite{lsz01}).
   a) and c) show the 3 and 1mm maps synthesized from 
   all uv data. b) and d) show maps derived from long $uv$\, spacings only.
   The synthesized beams (FWHM) are shown as grey ellipses at the lower 
   left corners.   
   Contour levels are in steps of (-2,2,4 to 22 by 3) times 1$\sigma$\,rms.
   a) $\sigma=0.6$\,mJy/ 1\asp2$\times$0\asp84 beam.
   b) $\sigma=0.75$\,mJy/ 1\asp0$\times$0\asp63 beam.  
   c) $\sigma=1.3$\,mJy/ 0\asp58$\times$0\asp39 beam.  
   d) $\sigma=1.4$\,mJy/ 0\asp50$\times$0\asp32 beam.  
}
\end{figure}
 
CB\,26 was observed with the Owens Valley Radio Observatory (OVRO) 
millimeter-wave array during 2000.  
We obtained wide-band continuum maps at 1.3 and 2.7\,mm and 
a $^{13}$CO(1--0) map with a spectral resolution of 0.17\,km/s. 
The angular resolution of the maps 
is $\sim$1\asp2$\times$0\asp8 at 2.7\,mm and 
$\sim$0\asp5$\times$0\asp3 at 1.3\,mm, respectively. 
The observations are described in more detail in \cite{lsz01}. 

The dust continuum maps 
show an elongated source with a position angle of 
$60\pm 3$\grd\ (from N to E; Fig. \ref{fig_mmcres}). 
The position and morphology of this source suggest that the emission 
originates from the circumstellar disk implied 
from the NIR data \cite{ste01}. 
Figure \ref{fig_jmh}b shows that the dust continuum emission 
matches well the extinction lane at the center of the NIR reflection nebula. 
At 3\,mm we derive a projected FWHM source size of 
$(160\pm10)$\,AU\,$\times$\,$<$\,60\,AU with no signature of an 
extended envelope. 
The 1\,mm images show a narrow lane  
and a small envelope extending perpendicular to the disk. 
The disk height remains unresolved at even the highest angular resolution 
of 0\asp3, confirming that it must be seen almost edge-on. 
The projected FWHM size of the 1.3\,mm disk is 
$(220\pm20)$\,AU\,$\times$\,$\le$\,20\,AU ($\frac{1}{2}$\ of the HPBW 
in that direction). The actual scale height of the inner disk where the 
mm emission arises is probably much smaller. 
The disk has a symmetric 20\grd\ warp outside $R\sim$100\,AU 
and is traced out to $\sim$200\,AU.
The total flux densities derived from the OVRO maps are 
$S_{\rm 2.7mm}=(22\pm5)$\,mJy and $S_{\rm 1.3mm}=(150\pm30)$\,mJy.
At 1.3\,mm, the IRAM 30-m single-dish flux of the 
unresolved component is completely recovered (160\,mJy; \cite{lh97}).
A simple decomposition yields 1.3\,mm flux densities for the disk and 
envelope of $S_{\nu}{\rm(disk)}=(80\pm20)$\,mJy and 
$S_{\nu}{\rm(env)}=(70\pm20)$\,mJy, respectively. 
Assuming $T_{\rm d}=30$\,K, $\kappa_{\rm d}{\rm (1.3mm)}=1$\,cm$^2$\,g$^{-1}$\ of dust 
and $M_{\rm H}/M_{\rm d}=110$\ we derive an envelope mass of 
$M_{\rm H}{\rm (env)}=(0.03\pm0.01)$\,M$_{\odot}$. The total mass of the more 
extended envelope seen in the single-dish maps is $(0.12\pm0.05)$\,M$_{\odot}$\ 
(\cite{hen01}, \cite{lsz01}). 
The disk appears to be considerably smaller in the 3\,mm dust emission than 
at 1\,mm. For the central $R\le 100$\,AU  
we derive a 1--3\,mm spectral index $\alpha=2\pm0.5$\ (envelope subtracted).  
Further out $\alpha$\ increases to reach 3.5-4 at 
$R\sim 180$\,AU. At larger radii the 3\,mm emission is not longer traced. 
This implies that most of the dust mass in the outer warped part of the disk 
is contained in classical, $\mu$m--mm size grains with a spectral index 
of the dust opacity of $\beta\sim 1.5-2$. 
The mm dust emission from the inner $\sim$200\,AU of the disk 
is highly optically thick. Therefore, no constraints can be made on 
$\beta$\ without better constraints on the temperature distribution 
and extensive modeling. 
Assuming that the 3\,mm emission is mostly optically thin, 
a lower limit to the disk mass can be derived.  
Adopting $\kappa_{\nu}{\rm (1.3mm)}=0.02$\,cm$^2$\,g$^{-1}$\ of ISM, 
$\beta=1$, $T_{\rm o}=1000$\,K at $r_{\rm o}=0.1$\,AU, $T\propto r^{-0.4}$, and 
surface density $\Sigma\propto r^{-1.5}$\ as 'typical' disk parameters, 
we derive $M_{\rm H}{\rm (disk)}\ge 0.01$\,M$_{\odot}$, a value which is 
typical for disks around T Tauri stars \cite{bec90}.

%This means that most of the dust mass in the disk, at least in its outer 
%parts, is still contained in $\mu$m--mm size grains and not yet in 
%cm size pebbles or planetesimals \cite{bec00}. 

%%%%%%%%%%%%%%%%%%%%%%%%%%%%%%%%%%%%%%%%

%\subsection{The rotation curve}

\begin{figure}[htb]
\begin{center}
\includegraphics[width=.5\textwidth]{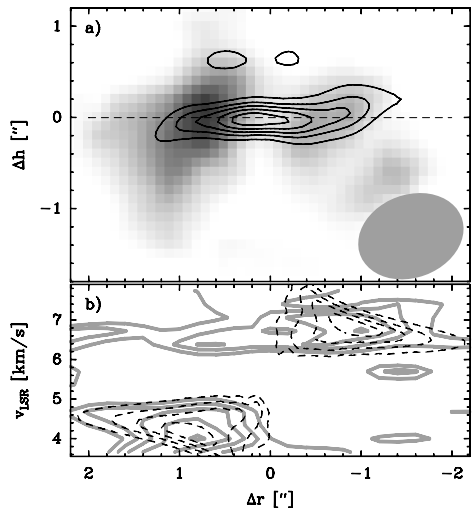}
\end{center}
\caption{\label{fig_cores}
   $^{13}$CO(1--0) emission from the CB\,26 disk: OVRO results.
   a) grey-scale: integrated intensity of the 
     $^{13}$CO(1--0) emission. The synthesized beam (FWHM) is 
     shown as grey ellipse in the lower right corner. Overlayed 
     are the contours of the 1.3\,mm dust continuum emission from the disk. 
     The image is rotated by 30\grd.~~~
   b) Position-velocity diagram of $^{13}$CO along the plane of the 
      disk (dashed line in a). Thick grey contours show the observed 
      velocity field (27, 45, 63, 81, 99\%\ of max). Dashed contours 
      show the modeled emission from a Keplerian disk around a 
      0.35\,M$_{\odot}$\ star.}
\end{figure}

Strong $^{13}$CO(1--0) emission ($T^{\ast}_{\rm R}\sim 20$\,K) 
was detected from the outer parts of the CB\,26 disk, but there seems 
to be a lack of emission from the central part (Fig. \ref{fig_cores}a). 
%The reason for this becomes clear from the velocity structure. 
The line is double-peaked
with the blue part coming from the N-E side and the red part 
from the S-W side of the disk. 
%The main peaks are separated 
%by $\sim$1\asp6 ($\sim$\,220\,AU) and $\sim$\,3\,km\,s$^{-1}$. 
Emission at the systemic velocity of the envelope 
$v_{\rm LSR}=5.5$\,km\,s$^{-1}$\ ($\Delta v\sim$\,0.6\,km\,s$^{-1}$) 
is self-absorbed and/or resolved out. 
Although the current spectral data cover only a small velocity range 
(3\,$\le$\,$v_{\rm LSR}$\,$\le8$\,km/s), they could 
be well-modeled by a Keplerian disk rotating around a 
(0.35$\pm$0.1)\,M$_{\odot}$\ star (Fig. \ref{fig_cores}b). 
The high-velocity line wings from the inner parts of the disk expected 
for a Keplerian disk are not recovered in the data because of the 
small band width. 
There is some 'forbidden' red-shifted emission from the 'blue side' 
of the disk which may be due to infall or outflow. 
A more detailed study of the kinematics of this disk-envelope 
system is subject to follow-up observations.

%%%%%%%%%%%%%%%%%%%%%%%%%%%%%%%%%%%%%%%%

\section{Conclusions}

\begin{figure}[htb]
\begin{center}
\includegraphics[width=.9\textwidth]{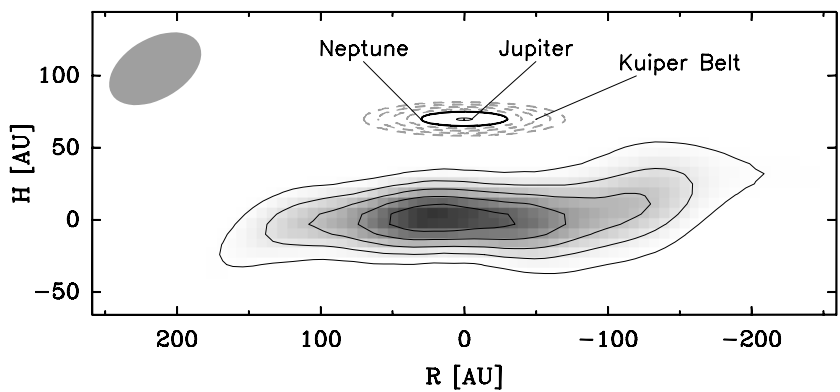}
\end{center}
\caption{\label{fig_solsys} 
        The CB\,26 disk compared to the solar system. 
        The grey scale image and contours show the 1.3\,mm 
        dust continuum emission from the CB\,26 disk as 
        observed with OVRO. For comparison, the solar system 
        is represented 
        by the orbits of Jupiter and Neptune and the 
        Kuiper Belt (40--70\,AU).}
\end{figure}

From OVRO millimeter interferometric observations we discovered 
and resolved the thermal dust continuum and $^{13}$CO line emission 
from the circumstellar disk in CB\,26 which was recently implied by 
high-resolution NIR polarimetric observations. 
The disk is seen edge-on as predicted by the NIR observations 
and matches well the extinction lane at the center of the bipolar 
NIR reflection nebula. 
It has a FWHM diameter of (220$\pm$20)\,AU and a maximum (traced) 
diameter of $\sim$400\,AU. 
The (projected) scale height remains unresolved ($h\le$\,20\,AU). 
For the first time, we directly detect a warp in a young circumstellar 
disk. The warp affects the disk outside $R> 100$\,AU. Like in many 
extra-galactic disks, such a warp could be caused by a small 
external disturbance like, e.g., an encounter with a nearby star. 
However, the actual cause of the warp is not known. 

The 1.3\,mm dust emission from the inner $R\sim 100$\,AU of the disk 
is highly optically thick, and only a lower limit to the total disk mass 
$M_{\rm H}{\rm (disk)}\ge 0.01$\,M$_{\odot}$\ can be derived from the 
3\,mm emission. 
Due to the high optical depths, no constraints on the dust opacity 
spectral index $\beta$\ and possible planetesimal growth in the inner 
undisturbed part of the disk can be drawn. 
The dust emission from the outer parts 
of the disk is optically thin and arises from small 'classical' grains. 
However, since the outer disk is obviously disturbed as indicated by 
the warp and has much lower particle densities, this does not exclude 
that particle growth in the inner disk has already taken place. 
The small disk envelope, which extends above and below 
the plane of the disk and may be related to an yet undetected outflow 
or a disk wind, contains $\sim (0.03\pm0.01)$\,M$_{\odot}$. 
Further $(0.12\pm0.05)$\,M$_{\odot}$\ are contained in a more extended 
envelope ($\sim 3000$\,AU in diameter) which extends mainly to the 
north-east and may represent the remnant, nearly dispersed 
protostellar core from which this system has formed \cite{hen01}. 

Strong $^{13}$CO emission shows that the disk must be still gas rich, 
that it is rotating, and has a kinetic temperature of order 30\,K 
outside $R\sim 100$\,AU. The rotation curve is consistent 
with Keplerian rotation around a (0.35$\pm$0.1)\,M$_{\odot}$\ star. 
With a total luminosity of $\ge 0.7$\,L$_{\odot}$, its large disk and 
its close connection with the parental cloud core, the CB\,26 system 
resembles an intermediate M type T\,Tauri star of age $< 10^5$\,yrs. 
The lack of a dense, centrally peaked cloud core with spectroscopic 
infall signatures as well as of a prominent molecular outflow 
indicates that the 
system has already passed its main accretion phase. Therefore, 
we conclude that this is a young protoplanetary disk.

If compared to the solar system today, the CB\,26 disk is $\sim2.5$\ 
times larger than the outer Kuiper Belt  (Fig. \ref{fig_solsys}).
However, the optically thick and undisturbed part 
inside the warp has approximately the size of the Kuiper Belt. 
The total mass in the CB\,26 disk (if all the gas is still there) 
is at least 10 times higher than the planetary mass in the solar system. 
However, the dust mass in the CB\,26 disk could be comparable 
or lower than the rocky mass in the solar system. 
These numbers depend critically on the dust properties which 
are not well-known.
Altogether, the CB\,26 system appears like a 2--3 times lower-mass 
equivalent to the early solar nebula at the verge of forming planetesimals 
and probably planets.

%%%%%%%%%%%%%%%%%%%%%%%%%%%%%%%%%%%%%%%%%%%%%%%%%%%%%%%%%%%%%%%%%%%

%INDEX%%%%%%%%%%%%%%%%%%%%%%%%%%%%%%%%%%%%%%%%%%%%%%%%%%%%%%%%%%%%%%%
% Please check with the editor of your book whether he plans to
% include a "mutual" subject index - if so, please code your entries
% in the standard syntax. For your own purposes you may print your
% "personal" index by using the following commands:
%
%\clearpage
%\addcontentsline{toc}{section}{Index}
%\flushbottom
%\printindex
%%%%%%%%%%%%%%%%%%%%%%%%%%%%%%%%%%%%%%%%%%%%%%%%%%%%%%%%%%%%%%%%%%%%%

\end{document}